\documentclass[showpacs, twocolumn]{revtex4-1}
\usepackage{graphicx}
\usepackage{amsmath}
\usepackage{appendix}

\begin{document}

\title{ Binary Bose-Einstein condensates in a disordered time-dependent potential }

\author{Karima Abbas$^{1,2}$ and Abdel\^{a}ali Boudjem\^{a}a$^{1,2}$}
\affiliation{$^1$ Department of Physics, Faculty of Exact Sciences and Informatics, Hassiba Benbouali University of Chlef, P.O. Box 78, 02000, Ouled-Fares, Chlef, Algeria. \\
$^2$Laboratory of Mechanics and Energy, Hassiba Benbouali University of Chlef, P.O. Box 78, 02000, Ouled-Fares, Chlef, Algeria.}
\email {a.boudjemaa@univ-chlef.dz}

\date{\today}

\begin{abstract}

We study the non-equilibrium evolution of  binary Bose-Einstein condensates in the presence of weak random potential with a Gaussian correlation function 
using the time-dependent  perturbation theory.
We apply this theory to construct a closed set of equations that highlight the role of the spectacular interplay between the disorder and the interspecies interactions
in the time evolution of the density induced by disorder in each component. It is found that this latter increases with time favoring localization of both species.
The time scale at which the theory remains valid depends on the respective system parameters. 
We show analytically and numerically that such a system supports a steady state that periodically changing  during its time propagation.
The obtained dynamical corrections indicate that disorder may transform the system into a stationary out-of-equilibrium states.
Understanding this time evolution is pivotal for the realization of Floquet condensates.

\end{abstract}

\maketitle

\section{Introduction}

Ultracold Bose mixtures of atomic gases offer unprecedented control tools, opening promising new avenues for the investigation of Bose-Einstein condensate (BEC)
in a disordered environment \cite{Boudj200, Boudj222}.
Disordered binay BECs present rich physics not encountered in a single component condensate due the intriguing interplay of quantum fluctuations induced by intra- and interspecies 
interactions and disorder effects.
Such dirty Bose mixtures could be regarded as a feasible simulator to analyze plethora of novel quantum phenomena such as supersolidity, and quantum glasses.

Dilute Bose-Bose gases in a weak disorder potential has recently undergone a resurgence due to their fascinating properties (see e.g \cite{Wehr,Nied,Xi,Boudj20}).
One of the most amazing features arising from the presence of the disorder in Bose mixture quantum gases is the modification of the miscibility criterion
and the dramatic phase separation between the two species \cite{Boudj200}.

The considerable interest in studying the dynamics of disordered BEC driven out-of-equilibrium
by slow (adiabatic) or sudden (quenched) changes to system parameters such as the scattering length has been boosted by 
remarkable advances in the tunability of ultracold atomic gases \cite{Bil, Roa,Laurent, Krin,Mel}. 
Non-equilibrium evolution of BEC offers the unique opportunity  to explore strongly correlated systems and transport in realistic physical systems.
Chen et {\it al}. \cite{Chen} have shown that  for a weakly non-equilibrium disordered Bose gas  under a quantum quench in the
interaction,  the disorder  can substantially destroy superfluidity more than the condensate leading to the so-called dynamical Bose glass.
Moreover, it has been revealed that externally controlled spatiotemporal periodic drive constitutes an excellent platform 
for creating novel nonequilibrium states of quantum matter \cite{Yuk1,Eck,Sing}.
Recent study demonstrates that the condensate deformation is a signature of the non-equilibrium feature of steady states of a Bose gas in a temporally controlled weak disorder \cite{Rad}.
Quite recently,  experimental realization of ultracold bosonic gases in  dynamic disorder with controlled correlation time have been reported in Ref.\cite{Nag}, 
where the microscopic origin of friction and dissipation has been well illustrated. 
There has been also an extensive amount of work addressing dynamics of bosons in disordered optical lattices (see e.g. \cite{Roa,Ander,Aubry,Choi,Ant,Gonz,White,Madro}).

Motivated by the above experimental and theoretical works, we investigate in this paper the non-equilibrium evolution of homogeneous Bose-Bose mixtures
subjected to a time-dependent disordered potential with a Gaussian correlation function.
To this end, we use the time-dependent perturbation theory.
The perturbation method has proven to be a very useful and powerful tool to capture the main features of both equilibrium and non-equilibrium disordered single BECs 
\cite{Boudj200, Rad, LSP, Lugan, Gaul, Krum, Nik, Boudj4}. 

We start with the equilibrium case and analyze some  ground-state and thermodynamic aspects of disordered mixtures. 
We look in particular at how the interplay of a correlated disorder (i.e. non-zero correlation length) and interspecies interactions affect the condensates deformation 
and their equations of state (EoS). 
Our analysis reveal that the condensates remain robust when the disorder correlation length is larger than the healing length 
regardless the strength of the interspecies interaction and of the disorder potential.
Comparison with our recent findings for Bose mixtures with white-noise potential \cite {Boudj200} 
and against other predictions such as the Huang-Meng predictions \cite{HM,Nag1} suggests the importance of the disorder correlation length in the localization process.

Furthermore, we study the dynamical properties of two BECs in a Gaussian disorder potential with time-periodic driving using the aforementioned time-dependent perturbation theory.
Periodically-driven quantum systems have gained tremendous interest recently owing to the possibility of {\it Floquet engineering} (see for review \cite{Holt}).
In the field of ultracold quantum gases, this concept could offer powerful techniques to reach novel phase transitions in condensed matter 
(see \cite{Holt, Holt1, Muk, Kon, Eck1, Lop, Jot, Aidel, Ken, Gold, Lell} and references therein).
We show that our theory predicts an oscillatory behavior of the condensates deformation during the time evolution.
It is found in addition that disorder drives the growth of the disorder fraction with time, reducing the condensed fraction in each species.
Therefore, the time scale for reaching a sizeable dynamical depletion in principle limits the validity of the present perturbation theory
Interestingly, a long time analysis predicts the existence of stationary states resembling to the Floquet states 
due to the combination of many-body effects, the drive frequency and disorder correlations.

The rest of the paper is organized as follows.
In section \ref{Equ}, we review the various formulas describing equilibrium Bose mixtures for arbitrary disorder potential.
We then restrict ourselves to symmetric mixture with Gaussian-correlated disorder. The results for asymmetric mixtures are shown in Appendix.
Section \ref{NEE} is the main section of this paper.  We introduce the time-dependent perturbative theory for disorder Bose mixtures.
We calculate the time-dependent condensate deformation due to the disorder. Many appealing issues are also discussed.
Finally, in section \ref{Conc}  we present conclusions and lessons learned.

\section{Equilibrium mixture} \label{Equ}

Consider weakly interacting homogeneous two-component BECs with equal masses, $m_1=m_2=m$, subjected to a weak random potential $U({\bf r})$, labeling the components by the index $j$.
At zero temperature, the system is described by the coupled Gross-Pitaevskii equations (GPE):
\begin{align} \label{GPE}
\mu_j \Phi_j = \left[- \frac{\displaystyle\hbar^2}{2m}\nabla^2 +U+g_j |\Phi_j |^2 + g_{12}  |{\Phi}_{\overline j} |^2 \right]\Phi_j,  
\end{align}
where $\Phi_j$ is wavefunction of each condensate, $\overline{j}=3-j$,  $\mu_j$ is the chemical potential of each condensate, 
$g_j=4\pi \hbar^2a_j/m$ and $g_{12}=g_{21}= 4\pi \hbar^2a_{12}/m$ with 
$a_j$ and $a_{12}$ being the intraspecies and the interspecies scattering lengths, respectively.
The disorder potential is assumed to have vanishing ensemble averages $\langle U({\bf r})\rangle=0$
and a finite correlation of the form $\langle U({\bf r}) U({\bf r'})\rangle=R ({\bf r}-{\bf r'})$.
In what follows, we restrict ourselves to the case of a Gaussian correlation with the Fourier transform \cite{Krum,Kob,Boudj1,Boudj2,Boudj3}
\begin{equation}  \label{Gauss}
R ({\bf k})=R_0 e^{- \sigma^2k^2/2}, 
\end{equation}
where $R_0$ is the disorder strength with dimension (energy)$^2$ $\times$ (length)$^3$ and $\sigma$ is the disorder correlation length. 

For weak disorder, Eq.(\ref{GPE}) can be solved using straightforward perturbation theory
in powers of $U$ using the expansion \cite{Boudj200} 
\begin{equation}  \label{Exps}
\Phi_j=\Phi_j^{(0)}+\Phi_j^{(1)} ({\bf r})+\Phi_j^{(2)} ({\bf r})+\cdots, \;\;\;\;\   j=1,2 
\end{equation}
where the index $i$ in the real valued functions $\Phi^{(i)}({\bf r})$ signals the $i$-th order contribution with respect to the disorder potential. 
They can be determined by inserting the perturbation series (\ref{Exps}) into the Eq.(\ref{GPE}) and by collecting the terms up to $U^2$.

\subsection {Glassy fraction} \label{GFF}

\begin{figure*}
\includegraphics[scale=0.62]{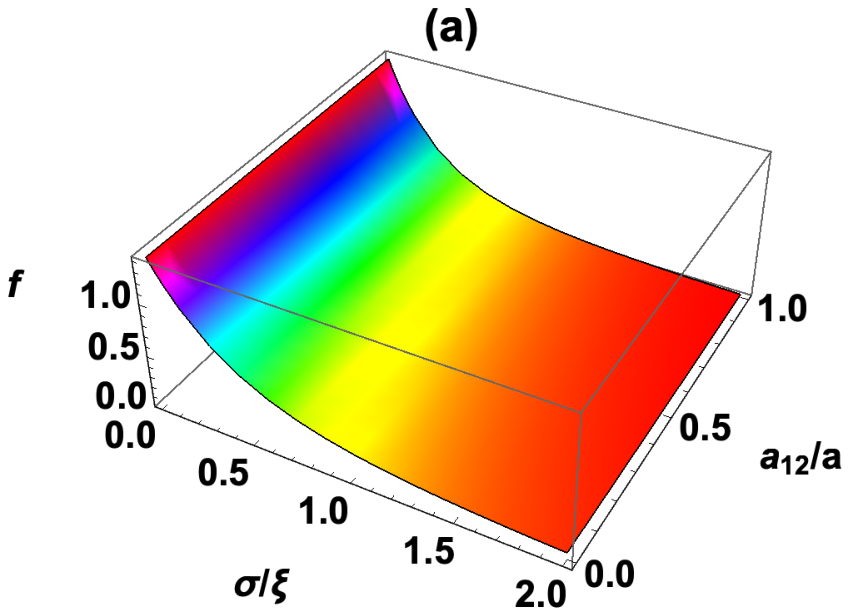}
\includegraphics[scale=0.62]{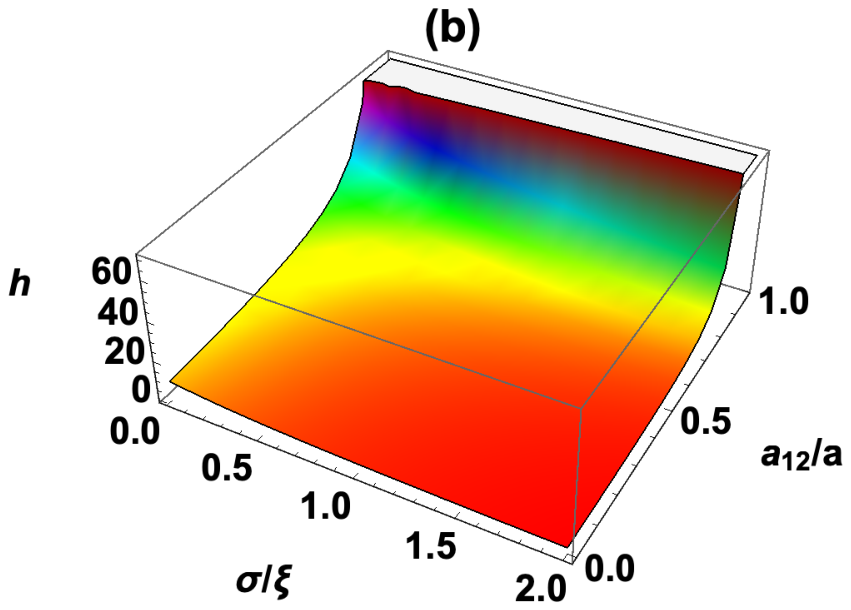}
\includegraphics[scale=0.62]{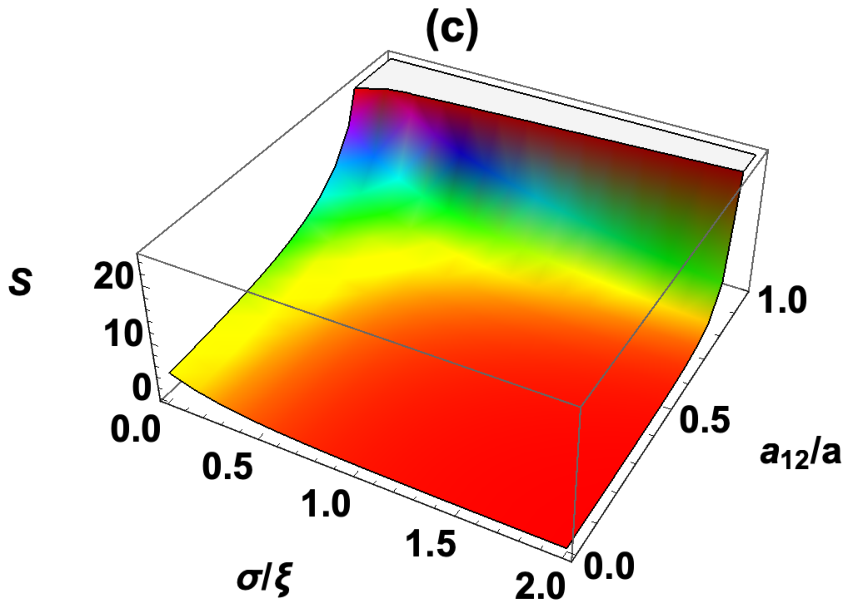}
\caption {Behavior of the disorder functions $f$ from Eq.(\ref{Funf}) (a), $h$  from Eq.(\ref{Funh})  (b), and $S$ from Eq.(\ref{FunS}) (c) as a function of $\sigma/\xi$ and $a_{12}/a$.}
\label{disG}
\end{figure*}

The condensate fluctuations  (or condensate deformation) due to  disorder, known as {\it glassy fraction}, can be given as the variance of the wavefunction
$n_{Rj}^{\text{eq}}= \langle \Phi_j^{(1)2} ({\bf r})  \rangle+\cdots$. Working in momentum space,  the glassy fraction reads \cite{Boudj200}:
\begin{align}\label{GF}
&n_{Rj}^{\text{eq}}= n_j\int\frac{d\bf k}{(2\pi)^3} R({\bf k}) \left [ \frac {E_k+2n_{\overline j} \big(g_{\overline j}-g_{12} \big)}  { {\cal E}_k } \right]^2, 
\end{align}
where ${\cal E}_k=\big(E_k+2g_j n_j \big) \big(E_k+2g_{\overline j} n_{\overline j}\big) - 4g_{12}^2 n_j n_{\overline j}$,
and $R ({\bf k})$ is the disorder correlation. 

Let us assume a symmetric mixture where $a_{1}=a_{2}=a $ and $n_1=n_2=n$.
Performing the integral (\ref{GF}), the glassy fraction turns out to be given: 
\begin{equation}\label{TDGF6}
 \frac{n_R^{\text{eq}}} {n}= \frac{\xi_+} {\ell_L}  f\left(\frac{\sigma}{\xi_+}\right), 
\end{equation}
where the disorder function reads
\begin{align} \label {Funf}
 f\left(\frac{\sigma}{\xi_+}\right) &= \sqrt{2}e^{\frac{\sigma ^2}{\xi_+ ^2}} \left(\frac{2 \sigma ^2}{\xi_+^2}+1\right) \text{erfc}\left(\frac{\sigma }{\xi_+}\right) 
-\frac{2\sqrt{2} \sigma } {\sqrt{\pi } \xi_+}, 
\end{align}
where $\text{erfc} (x)$ is the complementary error function, $\xi_+=\xi/\sqrt{\delta a_+}$ is the extended healing length for a symmetric mixture with short-range interactions,
$\xi=\hbar/\sqrt{2mng}$, $\delta a_+=1+a_{12}/a$, and $\ell_L= 4\pi \hbar^4/\left(m^2R_0\right)$ accounts for the Larkin length which is associated with the
pinning energy due to the disorder \cite {Rad,Natt,Flac}. 
For $\sigma \rightarrow 0$,  $f(0) =\sqrt{2}$, thus, the results of binary BECs with a weak delta-correlated disorder is recovered \cite{Boudj200}.
For $\sigma \rightarrow 0$ and $a_{12}=0$, one can reproduce the seminal Huang-Meng findings for a dirty single BEC \cite{HM}.

The behavior of the function (\ref{Funf}) is shown in Fig.\ref{disG}.a.
As expected, the  function $f$ is decreasing with the disorder correlation length $\sigma /\xi$ regardless of the strength of interspecies interactions
indicating that the condensate depletion due to the disorder effects is supressed for $\sigma \gg \xi$. 
One might explain this delocalization as the results of a screening of disorder by the interspecies interaction.
The same situation takes place in single dirty dipolar and nondipolar BECs \cite{Krum,Kob,Boudj1,Boudj4,Boudj5}.
For fixed  $\sigma$, $n_R^{\text{eq}}$ is decreasing with the interspecies interactions $a_{12}/a$.

The validity of the present perturbation theory requires to have the condensate depletion due to the disorder much smaller than the total density
$ n_R^{\text{eq}} \ll n$. This implies that to have a weak disorder potential, the following condition must be fulfilled
\begin{equation}\label{Vald}
\ell_L \gg \xi_+   f\left(\frac{\sigma}{\xi_+}\right).
\end{equation}
Equation (\ref{Vald})  is a natural extension of the result of \cite {Rad}.

\subsection{Equation of state} \label{EoS}

The EoS in second-order of the disorder strength is given by \cite{Boudj200}
\begin{align} \label {eos}
&\mu_j=g_j n_j+g_{12} n_{\overline j}+ \int\frac{d \bf k}{(2\pi)^{3}}\frac{R(k)}{ (g_j g_{\overline j}-g_{12}^{2}) {\cal E}_k^2 E_k} \\
&\times \Bigg\{ 4g_j^{2}g_{\overline j} n_j \left(E_k+g_j n_j\right) (E_k+2g_{\overline j} n_{\overline j})^{2} \nonumber\\
&+4g_j g_{\overline j} g_{12}n_{\overline j} \bigg[(E_k+g_{\overline j} n_{\overline j})(E_k+2g_j n_j)^2 \nonumber\\
&+E_k^{2} (E_k+2g_{\overline j} n_{\overline j} )\bigg]\bigg\} \nonumber.
\end{align}
In Eq.(\ref{eos}),  higher order terms in $g_{12}$ were omitted.

Upon calculating the integral (\ref{eos}), for the spacial case of $n_1=n_2=n$, and $a_1=a_2=a$, we find for the EoS 
\begin{equation} \label{eos1}
\mu=ng\Big[\delta a_{+}+\frac{\xi_-}{\ell_{L}} h\left(\delta a_{\pm},\frac{\sigma}{\xi\pm} \right)\Big],
\end{equation}
where the disorder functions read
\begin{widetext}
\begin{align} \label{Funh}
h \left(\delta a_{\pm},\frac{\sigma}{\xi_{\pm}}\right)=&\frac{\sqrt{\frac{2}{\pi}}}{ \delta a_{-}}\Bigg\{\left(\frac{\sigma}{\xi_{-}}\right)
\left[\frac{(\delta a_{+}-\delta a_{-})^{2}}{2\delta a_{+}\delta a_{-}}-2\right]+\frac{\sqrt{\pi\delta a_{-}}}{(\delta a_{+}-\delta a_{-})}\Big(A_{-}-A_{+}\Big)
+\frac{1}{\delta a_{+}}\Big[ 2\left(\frac{\sigma}{\xi_{-}}\right)(\delta a_{-}-\delta a_{+}) \\
&+\sqrt{\pi\delta a_{-}}\Big(B_{+}-B_{-}\Big)\Big] \Bigg\}, \nonumber
\end{align}
where 
$A_{\pm}=\frac{e^{(\sigma/\xi_{\pm})^{2}}}{(\delta a_{\pm})^{3/2}} \text{erfc}\left(\frac{\sigma}{\xi_{\pm}}\right)\left[7+\left(\frac{\sigma}{\xi_{\mp}}\right)^{2}-\left(\frac{\sigma}{\xi_{\pm}}\right)^{2}-9\delta a_{\pm}-\frac{(\delta a_{\mp}-\delta a_{\pm})^{2}}{4}\left(6+\left(\frac{\sigma}{\xi_{\mp}}\right)^{2}+5\left(\frac{\sigma}{\xi_{\pm}}\right)^{2}\right)\right]$,
$B_{\pm}=\sqrt{\delta a_{\pm}}\left[3+2\left(\frac{\sigma}{\xi\pm}\right)^{2}\right] e^{(\sigma/\xi_{\pm})^{2}} \text{erfc}\left(\frac{\sigma}{\xi_{\pm}}\right)$,
$\xi_{-} = \xi/\delta a_-$, and $\delta a_-=1-a_{12}/a$.
In the limit $\sigma \rightarrow 0$, one has:
\begin{equation} \nonumber
h_{\sigma\rightarrow 0}=\frac{2(\sqrt{\delta a_{-}}-\sqrt{\delta a_{+}})-3\left(\frac{a_{12}}{a}\right)^{2}(\sqrt{\delta a_{-}}-5\sqrt{\delta a_{+}})-12\left(\frac{a_{12}}{a}\right)^{3}(\sqrt{\delta a_{-}}+\sqrt{\delta a_{+}})+\left(\frac{a_{12}}{a}\right)(13\sqrt{\delta a_{-}}+\sqrt{\delta a_{+}})}{\sqrt{2}\sqrt{\delta a_{-}}\left(\frac{a_{12}}{a}\right)\left(1-\left(\frac{a_{12}}{a}\right)^{2}\right)^{3/2}}
\end{equation}
which is in agreement with our recent result for equilibrium Bose mixtures with uncorrelated disorder \cite{Boudj200}.
Evidently, Eq.(\ref{eos1}) shows that the EoS increases linearly with the disorder strength $R_0$.
\end{widetext}

Figure \ref{disG}.b shows that disorder function $h$ is lowering with $\sigma/\xi$ and rising with $a_{12}/a$.
This signals that the interplay of the disorder effects and interspecies interaction could modify the behavior of the EoS of Bose mixtures.

\subsection{Sound velocity} \label{MC}

Corrections to the sound velocity of each component due to the disorder fluctuations are given by \cite{Boudj2}:
\begin{equation} \label {SV}
c_{sj}^2= \frac{n_j}{m_j} \frac{\partial \mu_j}{\partial n_j}. 
\end{equation}
In the case of a balanced mixture $a_{1}=a_{2}=a $ and $n_1=n_2=n$, we find after a straightforward calculation:
\begin{equation}
\frac{c_{s}^{2}} {c_{s0}^2}=1+\frac{\xi_-}{\ell_{L}}S \left(\delta a_{\pm},\frac{\sigma}{\xi\pm}\right), 
\end{equation}
where $c_{s0}= \sqrt{ g n/m}$ is the zeroth-order sound velocity and the disorder function $ S(\delta a_{\pm}, \sigma/\xi\pm)$  is given as:
\begin{equation}\label{FunS}
S(\delta a_{\pm},\frac{\sigma}{\xi\pm})=\frac{1}{2} \left(h+2n\frac{\partial h}{\partial n}\right),
\end{equation}
which has practically the same behavior as the function $h$ as is displayed Fig.\ref{disG}.c.

\section{Non-equilibrium evolution} \label{NEE}

We consider two weakly interacting ultracold Bose gases, subjected to a weak random potential $ U({\bf r},t)=u({\bf r}) F(t)$, such that $ F(0)=0 $ and $ 0\leq F(t)\leq1$.
The system evolution at $t\geq0$ is described by the coupled time-dependent GP equations which can be written as :
\begin{align} \label{TDGP}
 i \hbar \frac{\partial}{\partial t} \Phi_{j}({\bf r},t) &= \bigg( -\frac{\hbar^{2}}{2m}\nabla^{2} + u({\bf r}) F(t) -\mu_{0j}\\
&+ g_{j}\vert\Phi_{j}({\bf r},t)\vert^{2} + g_{12}\vert\Phi_{\bar{j}}({\bf r},t)\vert^{2}\bigg)\Phi_{j}({\bf r},t). \nonumber
\end{align}
For sufficiently small $u({\bf r})$, the system can be treated perturbatively. Therefore, we can write the wavefunctions as:
\begin{equation} \label{Expst}
  \Phi_{j}({\bf r},t)= \Phi_{j}^{(0)}({\bf r})+\Phi_{j}^{(1)}({\bf r},t)+\Phi_{j}^{(2)}({\bf r},t)+\cdots, 
\end{equation}
where $ \Phi_{j}^{(0)}({\bf r})$ are the equilibrium solutions of Eq.(\ref{GPE}) at $t=0$, and $\Phi_{j}^{(\alpha)}(r,0)=0$ for  $\alpha\geq1$. 
The particle densities $n_{j}=\vert\Phi_{j}(r)^{(0)}\vert^{2}$ determine the chemical potentials of the system in its equilibrium ground-state: 
\begin{equation}
\mu_{0j}= g_{j}n_{j} + g_{12}n_{\bar{j}}.
\end{equation}
The condendate deformation due to the disorder potential, can be given by
 \begin{equation} \label{TDGF}
 n_{Rj}(t)=\langle\vert \Phi_{j}({\bf r},t)\vert^{2}\rangle-\vert\langle\Phi_{j}({\bf r},t)\rangle\vert^{2}.
 \end{equation}
 Using the perturbative expansion (\ref{Expst}) up to second-order, and assuming that the disorder have a vanishing ensemble averages the deformation of each BEC becomes:
 \begin{equation} \label{TDGF1}
 n_{Rj}(t)=\langle\vert \Phi_{j}^{(1)}({\bf r},t)\vert^{2}\rangle.
\end{equation} 

The first-order coupled equations follow from Eq.(\ref{TDGP}) read:
 \begin{align} \label{TDGP1}
 i \hbar \frac{\partial}{\partial t} \Phi_{j}^{(1)}({\bf r},t) &= \left( -\frac{\hbar^{2}}{2m}\nabla^{2}+g_{j}n_{j}\right)\Phi_{j}^{(1)}({\bf r},t) \\
 &+ g_{j}n_{j}\Phi_{j}^{(1)*}({\bf r},t) + \sqrt{n_{j}}u({\bf r})f(t) \nonumber\\
&+g_{12}\sqrt{n_{j}n_{\bar{j}}} \left[\Phi_{\bar{j}}^{(1)}({\bf r},t)+\Phi^{(1)*}_{\bar{j}}({\bf r},t)\right], \nonumber\\
- i \hbar \frac{\partial}{\partial t} \Phi_{j}^{(1)*}({\bf r},t) &= \left( -\frac{\hbar^{2}}{2m}\nabla^{2}+g_{j}n_{j}\right)\Phi_{j}^{(1)*}({\bf r},t) \\
&+ g_{j}n_{j}\Phi_{j}^{(1)}({\bf r},t) + \sqrt{n_{j}}u({\bf r})f(t) \nonumber \\
&+ g_{12}\sqrt{n_{j}n_{\bar{j}}} \left[\Phi_{\bar{j}}^{(1)}({\bf r},t)+\Phi^{(1)*}_{\bar{j}}({\bf r},t) \right].\nonumber
\end{align}
In Fourier space Eqs.(\ref{TDGP1}) turn out to be given as:
\begin{align}
&\left( \hbar\omega_{\bf k}+g_{j}n_{j}-i \hbar s\right)\Phi_{j}^{(1)}({\bf k},s)  + g_{j}n_{j}\Phi_{j}^{(1)*}({\bf k},s) \label{TDGP2} \\
& + g_{12}\sqrt{n_{j}n_{\bar{j}}} \left[\Phi_{\bar{j}}^{(1)}({\bf k},s)+\Phi^{(1)*}_{\bar{j}}({\bf k},s)\right]=- \sqrt{n_{j}}u({\bf k})f(s), \nonumber\\
& \left(\hbar\omega_{\bf k}+g_{j}n_{j}+i \hbar s\right)\Phi_{j}^{(1)*}({\bf k},s)  + g_{j}n_{j}\Phi_{j}^{(1)}({\bf k},s) \label{TDGP22}\\
&+ g_{12}\sqrt{n_{j}n_{\bar{j}}} \left[\Phi_{\bar{j}}^{(1)}({\bf k},s)+\Phi^{(1)*}_{\bar{j}}({\bf k},s) \right]=- \sqrt{n_{j}}u({\bf k})f(s),\nonumber
\end{align}
where $\hbar\omega_{\bf k}=E_k$ is the kinetic energy. 
Here we used the Laplace transform $F(s)=\int_{0}^{\infty}dt\,F(t)e^{ -st}$.  
The solution of Eqs.(\ref{TDGP2}) and (\ref{TDGP22}) reads
\begin{align}
\Phi^{(1)}_{j}({\bf k},s)&=-\sqrt{n_{j}}u({\bf k})f(s) \\
&\times \frac{(\omega_{\bf k}+i s) \left[2g_{12}n_{\bar{j}} \omega_{\bf k}-\hbar  (\Omega^{2}_{{\bf k}\bar{j}}+s^{2})\right]} 
{4g_{12}^{2}n_{j}n_{\bar{j}} \omega_{\bf k}^2-\hbar^2(\Omega^{2}_{{\bf k}\bar{j}}+s^{2})(\Omega^{2}_{{\bf k}j}+s^{2})}, \nonumber\\
\Phi^{(1)*}_{j}({\bf k},s)&=-\sqrt{n_{j}}u({\bf k})f(s) \\
&\times \frac{(\omega_{\bf k}-i s) \left[2g_{12}n_{\bar{j}}\omega_{\bf k}-\hbar (\Omega^{2}_{{\bf k}\bar{j}}+s^{2})\right]}{4g_{12}^{2}n_{j}n_{\bar{j}}
\omega_{\bf k}^2-\hbar ^2(\Omega^{2}_{{\bf k}\bar{j}}+s^{2})(\Omega^{2}_{{\bf k}j}+s^{2})}, \nonumber
\end{align}
where  $\hbar \Omega_{{\bf k}j}=\sqrt{\hbar\omega_{\bf k}(\hbar\omega_{\bf k}+2g_{j}n_{j})}$ is the standard dispersion relation for a single BEC.
Using the inverse Laplace transform we get:
\begin{equation}
  \begin{split}
  &\Phi^{(1)}_{j}({\bf k},t)=-\frac{\sqrt{n_{j}}}{\hbar}u({\bf k})\int_{0}^{t}dt^{'}{\cal K}_{j}({\bf k},t-t^{'})f(t^{'}),\\
  &\Phi^{(1)*}_{j}({\bf k},t)=-\frac{\sqrt{n_{j}}}{\hbar}u({\bf k})\int_{0}^{t}dt^{'} {\cal K}_{j}^{*}({\bf k},t-t^{'})f(t^{'}),\\
   \end{split}
\end{equation}
where
\begin{equation} \label{KK}
 { \cal K}_{j}(k,t)=\frac{{ \cal K}_{j-}(k,t)-{ \cal K}_{j+}(k,t)}{(\Omega_{kj+}^{2}-\Omega_{kj-}^{2})},
\end{equation}
and
\begin{align} \nonumber
{ \cal K}_{j\pm}(k,t)&=\Big[i\cos(\Omega_{kj\pm}t)+\frac{\omega_{kj}}{\Omega_{kj\pm}}\sin(\Omega_{kj\pm}t)\Big] \\
&\times\Big[\Omega^{2}_{k\bar{j}}-\Omega_{kj\pm}^{2}-\frac{2g_{12}n_{\bar{j}}\omega_{k\bar{j}}}{\hbar}\Big], \nonumber
\end{align}
where the Bogoliubov spectrum of two-component BECs  is given as (see e.g \cite{Boudj6} and references therein):
\begin{equation} \label {Bog}
\Omega_{k \pm}= \sqrt{\omega_k^2+ k^2 c_{s\pm}^2},
\end{equation}
with $c_{s\pm}^2= \left(c_{s1}^{(0)2}/2\right) \left[1 + \bar\mu \pm \sqrt{ (1-\bar\mu)^2 +4 \Delta ^{-1}\bar\mu }\right]$
being the sound velocities in the density ($c_{s-}$) and spin ($c_{s+}$) channels, $\bar{\mu}_j=n_{\overline j} g_{\overline j}/n_j g_j$, and $\Delta=g_j g_{\overline j}/g_{12}^2$.
In the limit $k \rightarrow 0$, the total dispersion is phonon-like $\Omega_{k \pm}= c_{s\pm} k$.

Using the fact that $\langle \Phi^{(1)*}_{j}({\bf k},t) \rangle=0$, the time-dependent disorder densities (\ref{TDGF1}) take the form:
\begin{equation} \label{TDGF3}
 n_{R_{j}}(t)=\frac{n_{j}}{\hbar^{2}}\int \frac{d{\bf k}}{(2\pi)^{3}} R({\bf k})\left\vert\int_{0}^{t}dt^{'}{\cal K}_{j}({\bf k},t-t^{'})f(t^{'})\right\vert^{2}.
\end{equation}
To exemplify outrightly the form of the density (\ref{TDGF3}),  let us suppose a time-periodic disorder potential which is abruptly switched on at time $t= 0$
\begin{equation}
F(t)=\frac{1}{2} \left[1-\cos (\omega t)\right],
\end{equation}
where $\omega$ is the external frequency.
For $t=(2j+1)\pi /\omega$ with $j$ being an integer, the function $F(t)$ reaches its maximum ($F(t)=1$). 
Therefore, the system earns a new stationary state as we shall see hereafter.

For a symmetric mixture where $a_{1}=a_{2}=a $ and $n_1=n_2=n$, the Boboliubov spectrum (\ref{Bog}) reduces to the following dimensionless form:
$\Omega_{k \pm}= \Omega_0 (k \xi)\sqrt{(k \xi)^2+2(1\pm a_{12}/a)}$, where $\Omega_0=ng/\hbar$ is the inverse characteristic mean-field time scale.
It is clearly seen that for $ a_{12}/a>1$,  the spectrum associated with the density channel $\Omega_{k -}$ becomes complex leading to destabilize the system.
While for $ a_{12}/a<-1$, the spectrum $\Omega_{k +}$ may become imaginary and thus, the mxiture would be destabilized.
Inserting $\Omega_{k \pm}$ and Eq.(\ref{KK}) into Eq.(\ref{TDGF3}), and keeping in mind that terms associated with $\Omega_{k -}$ cancel, one finds for $n_{R}(t)$:

\begin{figure}[htb] 
\includegraphics[scale=0.45]{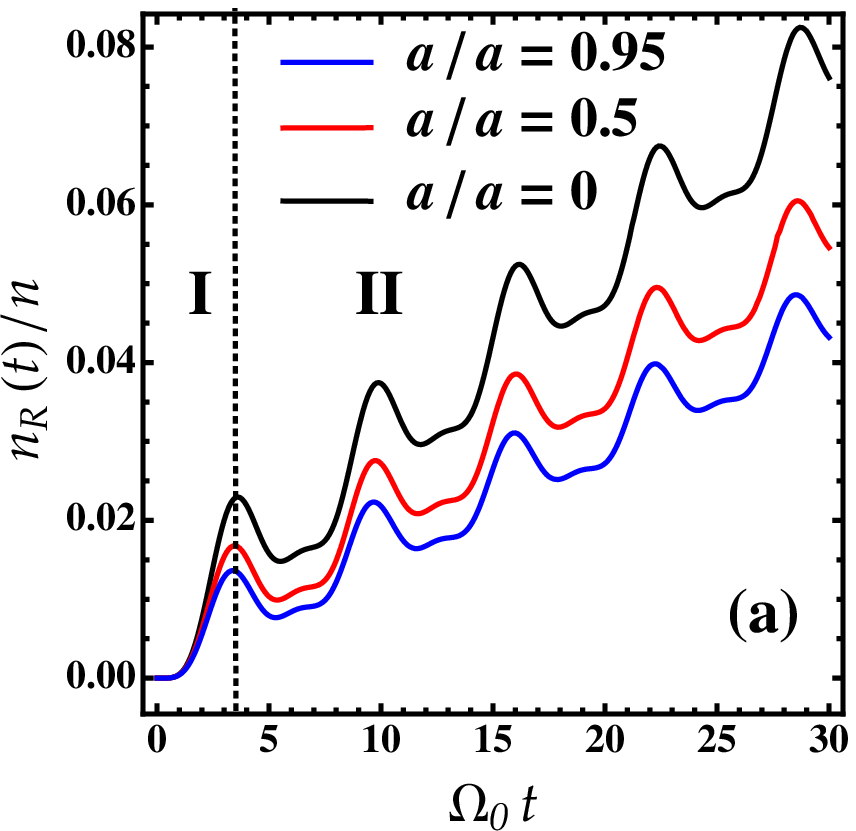}
\includegraphics[scale=0.46]{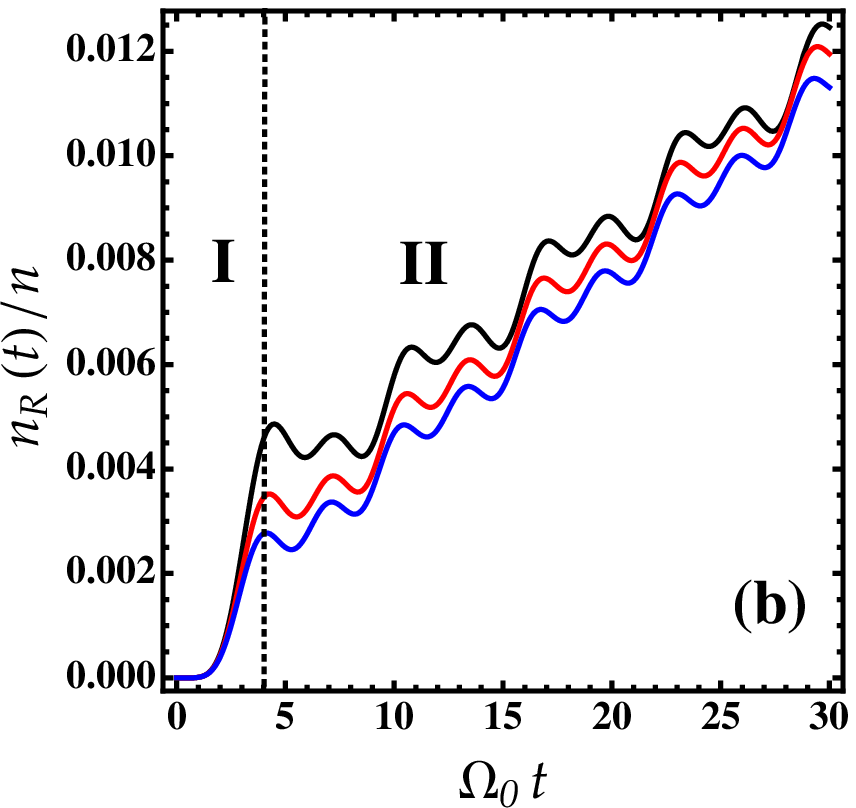}
\caption {(Color online) (a) Time evolution of the disorder fraction from Eq.(\ref{TDGF4}) for different values of $a_{12}/a$.
Parameters are : $a/a_0 = 95.44$, $n = 10^{21}$m$^{-3}$, $R'=0.5$, $\omega/\Omega_0=1$, (a) $\sigma/\xi=0.2$, and (b) $\sigma/\xi=2$.
Here $R'=R_0/g^2n$.
There are two different regimes: I and II  delineated by a vertical dotted line.}
\label{dis33}
\end{figure}

\begin{align} \label{TDGF4}
 n_{R}(t)=&\frac{n}{4\hbar^{2}}\int \frac{d{\bf k}}{(2\pi)^{3}}\frac{R({\bf k})}{\Omega_{\bf k+}^{4}\left(\Omega_{\bf k+}^{2}-\omega^{2}\right)^{2}} \\
&\times\Bigg\{\omega^{2}\Omega_{\bf k+}^{2} \left[\Omega_{\bf k+}\sin(\omega t)-\omega\sin(\Omega_{\bf k+}t) \right]^{2}\nonumber\\
&+\omega_{\bf k}^{2} \left[\omega^{2}-\Omega_{\bf k+}^{2}+\Omega_{\bf k+}^{2}\cos(\omega t)-\omega^{2}\cos(\Omega_{\bf k+}t)\right]^{2}\Bigg\}\nonumber.
\end{align}
This equation tells us that once the depletion due to disorder is known at $t=0$, both the condensed density $n-n_R$ and $n_R$  
can be calculated in a somehow simpler way at time $t >0$.

For a given $k$ and $\omega>0$, the density (\ref{TDGF4})  is peaked "resonant" at frequencies $\Omega_{k+}=\omega$ 
(i.e. when the external frequency nearly matches the eigenfrequencies). In terms of momenta this condition yields  
$$k=k_{\text{res}}= \frac{mc_s\sqrt{1+a_{12}/a}}{ \hbar\sqrt{-2+2\sqrt{1+\hbar^2\omega^2/m^2c_s^4 (1+a_{12}/a)^2}} }.$$
For $k \geq k_{\text{res}}$, the disorder depletion becomes very large, $n_{R}(t)/n \gg1$, indicating that the perturbation theory is no longer valid in such an unstable regime
despite the fact that the disorder is naively weak. Therefore, the intuitive stability criterion reads $k<k_{\text{res}}$.

\begin{figure*}[htb] 
\includegraphics[scale=0.52]{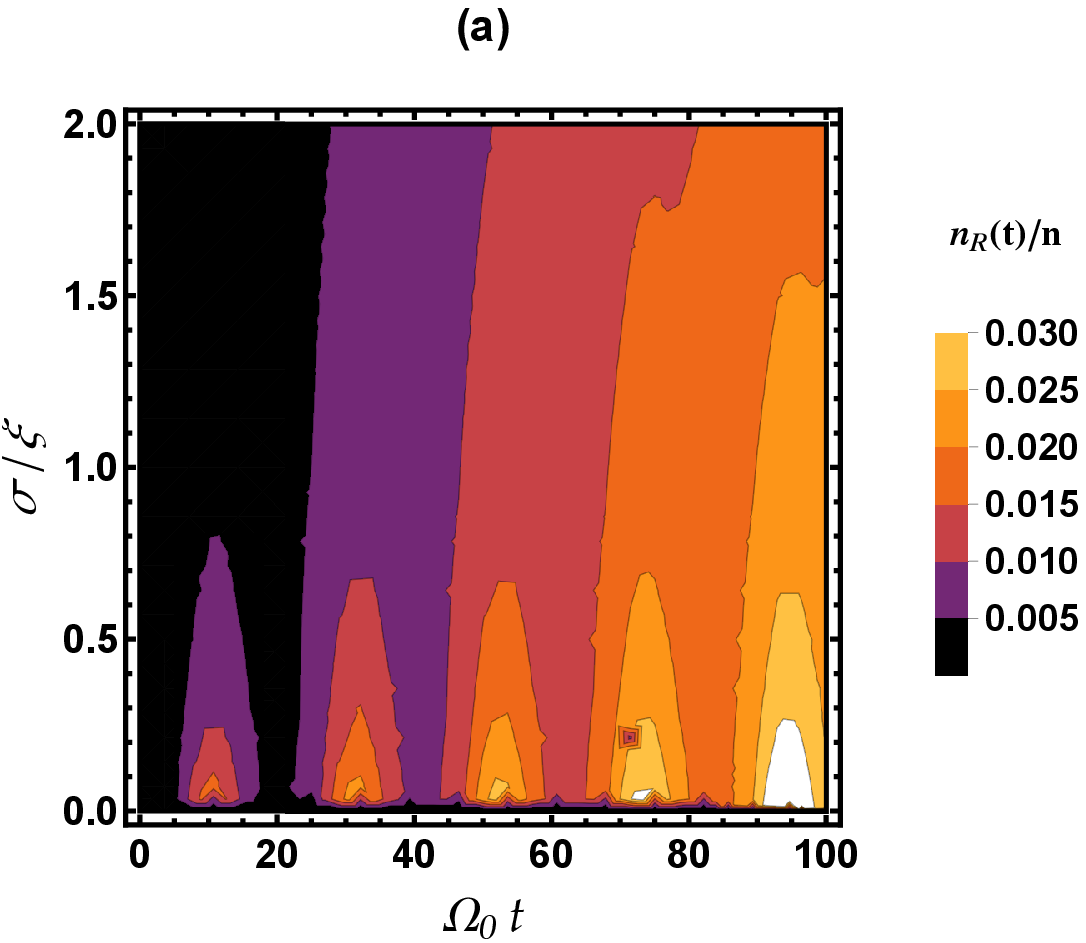}
\includegraphics[scale=0.52]{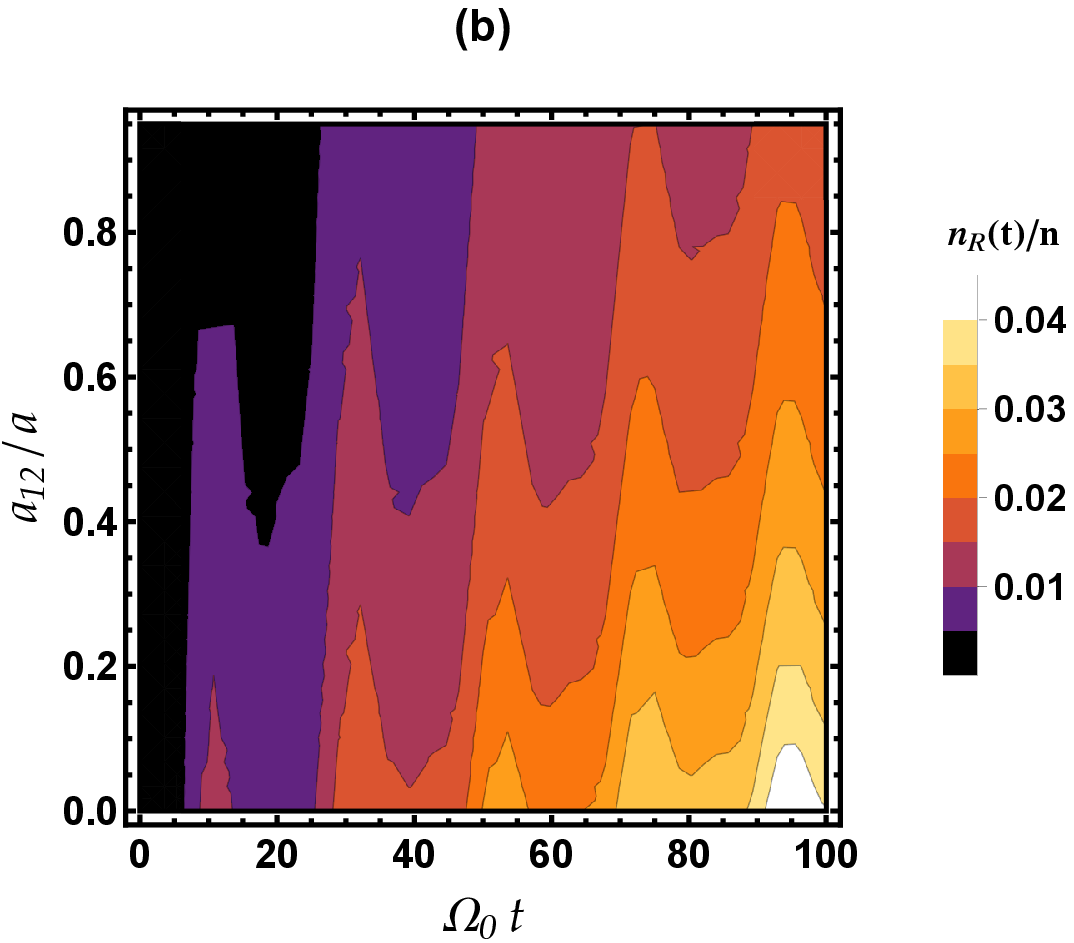}
\includegraphics[scale=0.52]{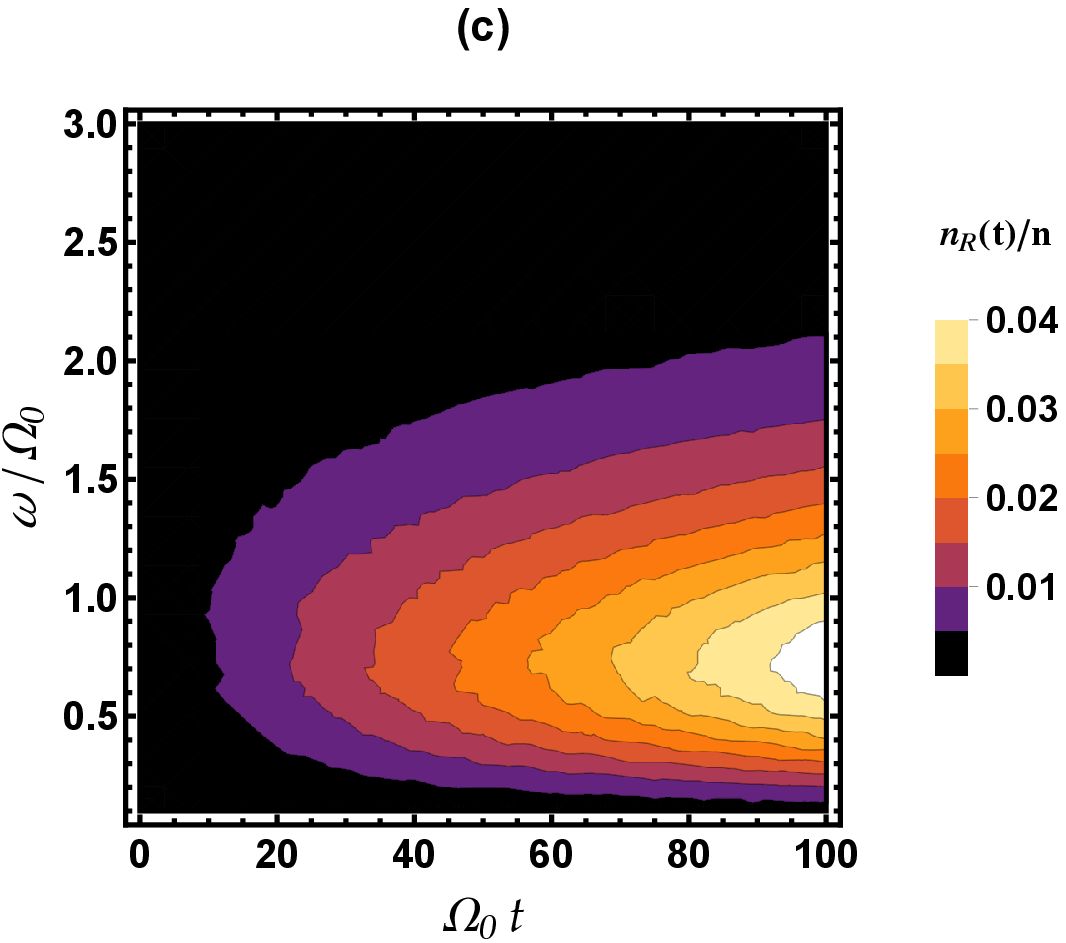}
\caption {(Color online)  (a) Disorder fraction $n_R(t)/n$ from Eq.(\ref{TDGF4}) as a function of $\Omega_0 t$ and $\sigma/\xi$ for $a_{12}/a=0.5$ and $\omega/\Omega_0=0.3$.
(b) $n_R(t)/n$ as a function of $\Omega_0 t$ and $a_{12}/a$ for $\sigma/\xi=0.5$ and $\omega/\Omega_0=0.3$. 
(c) $n_R(t)/n$ as a function of $\Omega_0 t$ and $\omega/\Omega_0$ for $\sigma/\xi=1.8$ and $a_{12}/a=0.5$.
Parameters are the same as in Fig.\ref{dis33}.}
\label{dis44}
\end{figure*}

In order to substantiate the relevance of the above time-dependent perturbative mean-field approach for laboratory experiments, 
we consider the ${}^{87}$Rb-${}^{87}$Rb mixture in two different internal states, but our theory can be readily generalized to other mixtures.
The scattering lengths and the densities are chosen to $a_1=a_2=a= 95.44\,a_0$ \cite{Ego} with $a_0$ being the Bohr radius, and 
$n_1=n_2=n=10^{21}$ m$^{-3}$, respectively, which are sufficient to ensure that the system meets the requirement of weakly interacting gas, $\sqrt{na^3}\simeq 10^{-2}\ll1$.
The interspcies scattering length $a_{12}$ which can be adjusted via Feshbach resonance is selected in such a way that the phase-separated condition is fulfilled throughout the dynamics.
The disorder strength is fixed to be $R'=0.5$ which gives $n_R/n \lesssim 1\%$, ensuring the sufficient criterion for the weak disorder regime.
We employ various disorder driving frequencies and correlation lengths. 

The numerical solutions of Eq.(\ref{TDGF4}) is shown in Fig.\ref{dis33}.
One can clearly identify two phases of evolution:
In phase I, $\Omega_0 t\lesssim 4$, although the density $n_R (t)$ is somehow low,
the dynamics follows an exponential growth due to the considerable effect of both quasiparticles and disorder onto the two BECs.
In region II, $\Omega_0 t > 4$, we observe that as the disorder evolves in time, the glassy fraction increases and exhibits sinusoidal oscillations
signaling that the system being completely depleted at long time.
This can be attributed to the motion of the Gaussian disorder which may create elementary excitations (i.e. the Bogoliubov phonons at low momenta/free particle in the high-energy regime)  
leading to enhance the disorder depletion. Therefore, whatever the strength and the frequency of the disordered potential, the condensates are localized 
since $n_R (t)$ extends to infinity at long times. 
In this region, the dynamics slightly slows down without any saturation and it presumably follows another growth law.
It is worth noticing that a very similar change in behavior has been observed in the dynamics of periodically-driven BEC in a shaken 1D lattice \cite{Lell}.

It is clearly visible that in the case of $\sigma <\xi$, the time-dependent glassy fraction is increasing with decreasing the interspecies interactions (see Fig.\ref{dis33}.a).
Whereas for $\sigma >\xi$, $n_R(t)$  varies with small oscillations and remains almost insensitive to $a_{12}/a$ which
means that the disorder density is protected against interspecies interactions effects during its time evolution (see Fig.\ref{dis33}.b).
Furthermore, as the healing length $\xi$ decreases, the chemical potential rises and the density of two BECs $n$ is growed.
In this situation  the low-lying excitations is poorly affected by the disorder time evolution, gives rise to lower the glassy fraction.

In Fig.\ref{dis44} we present the time evolution of the disorder fraction as a function of the relevant parameters.
We see that at short times $\Omega_0 t\lesssim 4$ the atoms are almost delocalized i.e. $n_R(t)$ is vanishingly small.
Whereas, $n_R(t)$ is increasing as the time goes on which implies a possibility for reducing the condensed fraction whatever the values of 
$\sigma /\xi$, $\omega/\Omega_0$, and $a_{12}/a$.
The dynamics slows down significantly for small external frequencies $\omega <\Omega_0$,  relatively strong interspecies interactions $a_{12} \gtrsim 0.5 a$ and for 
large disorder correlation $\sigma>\xi$ as is displayed in Fig.\ref{dis44}.a, b and c.
Figure \ref{dis44}.a and c depicts also that the oscillation strength of the condensate deformation strongly depends on $\sigma /\xi$ and $\omega/\Omega_0$.

Most noteworthy, for fixed $\sigma>\xi$ and varying the ratio $\omega/\Omega_0$, the density $n_{R}(t)$ remains practically constant in time for $\Omega_0 t \gtrsim 10$, 
apart from tenuous wigglings  (see Fig.\ref{dis44}.c) which could be an indicator of the existence of stationary Floquet condensates \cite{Holt,Holt1}.
Frankly speaking, a qualitative analysis of  these Floquet states in the presence of such periodic perturbations requires further thoughts.





\begin{figure}[htb] 
\includegraphics[scale=0.8]{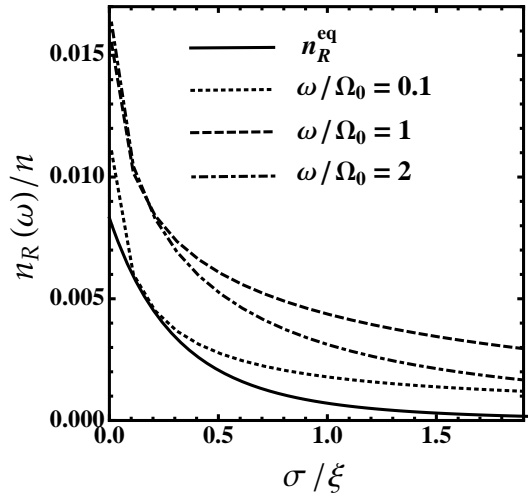}
\caption {Disorder fraction from Eq.(\ref{TDGF5}) as a function of $\sigma/\xi$.
Parameters are : $a/a_0 = 95.44$, $n = 10^{21}$m$^{-3}$, $R'=0.5$ and  $a_{12}/a=1$.}
\label{dis55}
\end{figure}

In the long-time limit $t=(2j+1)\pi /\omega$,  Eq.(\ref{TDGF4}) reduces to
\begin{align} \label{TDGF5}
 n_{R}(\omega)=&\frac{n}{4\hbar^{2}}\int \frac{d{\bf k}}{(2\pi)^{3}}\frac{R({\bf k})}{\Omega_{\bf k+}^{4}\left(\Omega_{\bf k+}^{2}-\omega^{2}\right)^{2}} \\
&\times\Bigg\{ 4\omega_{\bf k}^2 \left[ \omega^2\sin^2\left(\frac{(2j+1)\pi\Omega_{\bf k+}} {2\omega}\right)  -\Omega_{\bf k+}^{2} \right]^{2} \nonumber\\             
&+\Omega_{\bf k+}^{2} \omega^4 \sin^2\left(\frac{(2j+1)\pi\Omega_{\bf k+}} {\omega}\right)\Bigg\}\nonumber.
\end{align}
This equation shows that the time-dependence cancels meaning that the state hardly moves over such a time interval.
The leading term in Eq.(\ref{TDGF5}) represents the equilibrium disorder fluctuations,
while the subleading terms are of dynamical origin  revealing the non-equilibrium feature of the mixture as is shown in Fig.\ref{dis55} (dotted and dashed lines).
It is clear that the stationary depletion due to the  disorder diverges from the equilibrium one in particular for a small correlation length regardless of the values of 
the interspecies interactions.

On the other hand, after the adiabatic introduction of the disorder followed by the adiabatic switch-off i.e. $t=\pi /\omega$ and for $\omega \rightarrow 0$,
the stationary  depletion  (\ref{TDGF5}) becomes close to the equilibrium state, namely 
$n_{R}=(n/8\pi^3\hbar^2 ) \int d {\bf k} [R({\bf k}) \omega_k^2/\Omega_{k+}^4 ]= n_{R}^{\text{eq}}$ as is depicted in Fig.\ref{dis55} (see dotted and solid lines).
Conversly for  $\omega \rightarrow \infty $,  one has from Eq.(\ref{TDGF5}),
$n_{R}(\omega)=\pi^{3/2} (\xi_+/\ell_L)  (\Omega_0/4\omega)^2 (\xi_+/\sigma)^3$.
Such a condensate deformation is a signature of the nonequilibrium property of steady states of BEC in a time-dependent disorder potential.

\section{Conclusions} \label{Conc}

We investigated the non-equilibrium evolution of  binary BECs subjected to time-dependent random potential with a Gaussian correlation function. 
To this end, we applied the time-dependent perturbative mean-field theory which allows us to reveal the spectacular interplay 
of the disorder and the interspecies interactions in the non-equilibrium regime.
The theory assumes weak interactions and weak disorder, hence it remains valid provided the depletion remains small throughout the full subsequent dynamics.

We first shed new light on our understanding of equilibrium process and also established some useful formulas for the glassy fraction and the EoS 
for both symmetric and asymmetric (see Appendix) Bose mixtures. 
We pointed out that for a large disorder correlation length, the localization of bosons is suppressed owing to the screening of the random potential by the interaction.

In the case of a Gaussian disorder potential with time-periodic driving, we showed that the complex combination of atomic interactions, the disorder potential, 
and time-periodic perturbations may uncover new phenomena in dirty Bose mixtures.
The disorder fluctuations grow with time and exhibits an oscillating character, its magnitude strongly depends on the system parameters. 
To date, there is no experimental work confirming this oscillating character of the condensate deformation.
We found that at short time such an evolution follows an exponential growth law while at larger times where the theory fails, the dynamics is characterized by another law.
Among the main results emerging from our study is that even though the disorder is naively  weak, 
it could have dramatic effects on the localization of atoms during the time evolution of the system. 
The present analysis revealed also the occurence of a stationary state due to the crucial role played by the drive frequency in the limit of large disorder correlation length. 
We conjecture the existence of specific parameters enabling one to transform dynamic BECs into Floquet condensates.

Our time-dependent results can readily be extended to the case of trapped mixtures under the assumption that the disorder correlation length should be 
much smaller than the spatial extent of the atomic cloud \cite{Rad,Boudj1}.
We believe that our predictions open up new perspectives for an experimental demonstration of the peculiar interplay between the disorder and 
interaction in the non-equilibrium regime.


\section*{Acknowledgments}
We acknowledge  Axel Pelster for fruitful discussions and useful comments about the paper.

\newpage

\begin{widetext}

\section*{Appendix}
\subsection*{Asymmetric mixtures}
In this appendix we extend our results to the case of asymmetric disordered Bose mixtures.

The time-independent glassy density can be obtained from Eq.(\ref{GF})
\begin{equation}\label{GF1}
n_{Rj}^{\text{eq}} = n_{j\text{HM}} f_j(\Delta, \sigma_{j}),
\end{equation}
where $n_{j\text{HM}} =2 \pi R_j' n_j\sqrt{ n_j a_j^3/\pi}$ is the HM result for equilibrium deformation of each BEC \cite{Boudj200}  with
$R_j'=R_0/g_j^2n_j$ being a dimensionless disorder strength, $ \sigma_{j}=\sigma/\xi_j$, $\Delta=g_j g_{\overline j}/g_{12}^2$, 
and $\xi_j=\hbar/\sqrt{mn_jg_j}$ is the healing length of each component.
The disorder functions of the glassy fraction $f_j(\Delta, \sigma_{j})$ of Eq.(\ref{GF1}) can be written as:
\begin{align}
&f_{j}(\Delta, \sigma_{j})=\frac{\sqrt{2/\pi} }{\sqrt{\beta_{j}\alpha_{j}}(\alpha_{j}-\beta_{j})^{3}}\Bigg\{-2\sigma_{j}\sqrt{\alpha_{j}\beta_{j}}(\alpha_{j}-\beta_{j})\Big[2\nu_{j}^{2}+\alpha_{j}^{2}+\beta_{j}^{2}-2\nu_{j}(\alpha_{j}+\beta_{j})\Big]+\sqrt{\pi}e^{\sigma_{j}^{2}\beta_{j}}\Big[3\nu_{j}^{2}\beta_{j}\sqrt{\alpha_{j}}\\
&+\sqrt{\beta_{j}}(\alpha_{j}\beta_{j})^{3/2}(5-4\sigma_{j}^{2}\nu_{j})+\sqrt{\alpha_{j}}[\nu_{j}^{2}\alpha_{j}+2\nu_{j}\beta_{j}\alpha_{j}(-3+\sigma_{j}^{2}\nu_{j})-2\nu_{j}\beta_{j}^{2}(1+\sigma_{j}^{2}\nu_{j})+\beta_{j}^{3}(-1+2\sigma_{j}^{2}(2\nu_{j}+\alpha_{j}))-2\sigma_{j}^{2}\beta_{j}^{4}]\Big]\nonumber\\
&+\sqrt{\pi}e^{\sigma_{j}^{2}\beta_{j}}\sqrt{\alpha_{j}}(\beta_{j}-\nu_{j})\Big[\nu_{j}[\alpha_{j}+(3+2\sigma_{j}^{2}(\alpha_{j}-\beta_{j}))\beta_{j}]+\beta_{j}[-5\alpha_{j}+\beta_{j}-2\sigma_{j}^{2}\beta_{j}(\alpha_{j}-\beta_{j})]\Big] \bigg[1-\text{erfc}\left(\sqrt{\sigma_{j}^{2}\beta_{j}}\right)\bigg] \nonumber\\
&+\sqrt{\pi}e^{\sigma_{j}^{2}\alpha_{j}}\sqrt{\beta_{j}}(\alpha_{j}-\nu_{j})\Big[\alpha_{j}[\alpha_{j}+2\sigma_{j}^{2}\alpha_{j}(\alpha_{j}-\beta_{j})-5\beta_{j}]+\nu_{j}[\beta_{j}+\alpha_{j}(3-2\sigma_{j}^{2}(\alpha_{j}-\beta_{j}))]\Big] \text{erfc}\left(\sqrt{\sigma_{j}^{2}\alpha_{j}}\right)\Bigg\}, \nonumber
\end{align}
where $\text{erfc} (x)$ is the complementary error function,
$\alpha_j= 1+\bar{\mu}_j+\sqrt{(1+\bar{\mu}_{j})^{2}-4\bar{\mu}_j\left(\Delta-1\right)/\Delta}$,
$\beta_j=1+\bar{\mu}_j-\sqrt{(1+\bar{\mu}_{j})^{2}-4\bar{\mu}_j\left(\Delta-1\right)/\Delta}$, and
$\nu_j=2\bar{\mu}_j \left(1- g_{12}/g_{\overline j}\right)$
For $\sigma\rightarrow 0 $, the disorder functions $f_j$ tend to the delta-function obtained in our recent work \cite{Boudj200}. 
In the regime $\sigma\rightarrow 0 $ and $\Delta \rightarrow  \infty$ (or $g_{12} \rightarrow 0$, equivalently),  one has 
$f_j (\infty,0)=1$. This immediately leads to reproduce the Huang and Meng result \cite{HM} for the single component disorder fraction. 
If $\sigma\rightarrow 0 $ and  $\Delta \rightarrow 1$ (or $g_{12} \rightarrow \sqrt{g_1g_2}$, i.e in the vicinity of the phase separation), 
the functions $f_j(1,0)$ are diverging and become complex for $\Delta <1$.  

For our simulations we consider the above ${}^{87}$Rb-${}^{87}$Rb mixture in two different internal states with the following parameters:
$a_1= 100.4\,a_0$, $a_2= 95.44\,a_0$ \cite{Ego}, and $n_1=n_2=10^{21}$ m$^{-3}$. 

The behavior of the disorder functions $f_j$ versus of the interspecies interactions and the disorder correlation length is displayed in Fig. \ref{dis}.
We see that for $\sigma<\xi_j $, the effects of the disorder fluctuations are important. 
Whereas they vanish in the case of $\sigma>\xi_j $ for any value of interspecies interactions leading to arrest the localization in both species.
For fixed $\sigma<\xi_j $, the disorder functions $f_j$ rise with $a_{12}$ and diverge near the vicinity of the transition between the miscible and immiscible phases ($a_{12}\simeq98 a_0$)
\cite{Boudj200}.

One should stress that in the case of an asymmetric mixture, the criterion validity of the pertubation theory can be established from Eq.(\ref{GF1}).
In such a situation, there exists always a critical disorder strength $ R'_c$ above which a transition from miscible to immiscible phase occurs \cite{Boudj200}.

The chemical potential of each component can be obtained from integral (\ref{eos}), which gives
\begin{equation}\label{GF2}
\mu_j=g_j n_j+g_{12}n_{\overline j}+ g_j n_{j\text {HM}}  h_j(\Delta, \sigma_{j}),
\end{equation}
The disorder functions $h_j(\Delta, \sigma_{j})$ associated with the EoS read:
\begin{equation}
h_j(\Delta, \sigma_{j})=\left(\frac{\Delta}{\Delta-1}\right)\left(H_{1j}(\Delta, \sigma_{j})+\frac{g_{12} n_{\overline j}} {n_jg_j}  H_{2j}(\Delta, \sigma_{j})\right),
\end{equation}
where
\begin{align} \nonumber
&H_{1j}(\Delta, \sigma_{j})=\frac{4\sqrt{2/\pi} }{(\alpha_{j}-\beta_{j})^{3}}\Bigg\{\frac{2\sigma_{j}(\alpha_{j}-\beta_{j})}{\alpha_{j}\beta_{j}}\Big[-\alpha_{j}\beta_{j}[8\bar{\mu}_j^{2}+\alpha_{j}^{2}+\beta_{j}^{2}-4\bar{\mu}_j(\alpha_{j}+\beta_{j})]-8\bar{\mu}_j\alpha_{j}\beta_{j}+(\alpha_{j}+\beta_{j})(4\bar{\mu}_j^{2}+\alpha_{j}\beta_{j})\Big]\\
&+\frac{\sqrt{\pi}}{\alpha_{j}^{3/2}}e^{\sigma_{j}^{2}\alpha_{j}}(\alpha_{j}-2\bar{\mu}_j)\Big[2\bar{\mu}_j[\alpha_{j}(-5+2\sigma_{j}^{2}(\alpha_{j}-\beta_{j}))+\beta_{j}]+\alpha_{j}[\alpha_{j}-2\sigma_{j}^{2}\alpha_{j}(\alpha_{j}-\beta_{j})+3\beta_{j}] \nonumber\\
&+\alpha_{j}[\alpha_{j}(\alpha_{j}+2\sigma_{j}^{2}\alpha_{j}(\alpha_{j}-\beta_{j})-5\beta_{j})+2\bar{\mu}_j(\beta_{j}+\alpha_{j}(3-2\sigma_{j}^{2}(\alpha_{j}-\beta_{j})))]\Big]
\text {erfc} \left(\sqrt{\sigma_{j}^{2}\alpha_{j}}\right) \nonumber\\
&-\frac{\sqrt{\pi}}{\beta_{j}^{3/2}}e^{\sigma_{j}^{2}\beta_{j}}(\beta_{j}-2\bar{\mu}_j)\Big[2\bar{\mu}_j\alpha_{j}+\beta_{j}[2\bar{\mu}_j\alpha_{j}+3\alpha_{j}-2\bar{\mu}_j(5+2\sigma^{2}_{j}\alpha_{j})]+\beta_{j}^{2}[1+6\bar{\mu}_j +4\bar{\mu}_j\sigma_{j}^{2}+\alpha_{j}(-5+2\sigma_{j}^{2}(1+2\bar{\mu}_j))]\nonumber\\
&-\beta_{j}^{3}[-1+2\sigma_{j}^{2}(1+2\bar{\mu}_j+\alpha_{j})]+2\sigma_{j}^{2}\beta_{j}^{4}\Big]
\text {erfc} \left(\sqrt{\sigma_{j}^{2}\beta_{j}} \right)\Bigg\},\nonumber
\end{align}
and
\begin{align} \nonumber
&H_{2j}(\Delta, \sigma_{j})=\frac{2\sqrt{2/\pi}}{(\alpha_{j}\beta_{j})^{3/2}(\alpha_{j}-\beta_{j})^{3}}\Bigg\{-2\sigma_{j}\sqrt{\alpha_{j}\beta_{j}}(\alpha_{j}-\beta_{j})\Big[2\alpha_{j}\beta_{j}[8+\alpha_{j}^{2}+\beta_{j}^{2}-4(\alpha_{j}+\beta_{j})]\\
&-2\bar{\mu}_{j}[-8\alpha_{j}\beta_{j}+(\alpha_{j}+\beta_{j})(4+\alpha_{j}\beta_{j})]\Big]+\sqrt{\pi}e^{\sigma_{j}^{2}\alpha_{j}}\beta_{j}^{3/2}(\alpha_{j}-2) 
\Big[\alpha_{j}[4\sigma_{j}^{2}\alpha_{j}^{3}+6\beta_{j}\bar{\mu}_{j}-2\alpha_{j}^{2} (-1+2\sigma_{j}^{2}(\beta_{j}+\bar{\mu}_{j})) \nonumber\\
&+\alpha_{j}(2\beta_{j}(-5+2\sigma_{j}^{2}\bar{\mu}_{j})+2\bar{\mu}_{j})]+2[-4\sigma_{j}^{2}\alpha_{j}^{3}+2\beta_{j}\bar{\mu}_{j}+2\alpha_{j}^{2}(3+2\sigma_{j}^{2}(\beta_{j}+\bar{\mu}_{j}))
+2\alpha_{j}(\beta_{j}-5\bar{\mu}_{j}-2\sigma_{j}^{2}\beta_{j}\bar{\mu}_{j})]\Big]\text {erfc} \left(\sqrt{\sigma_{j}^{2}\alpha_{j}}\right) \nonumber\\
&+\sqrt{\pi}e^{\sigma_{j}^{2}\beta_{j}}\alpha_{j}^{3/2}(2-\beta_{j})\Big[4\beta_{j}[\alpha_{j}+2\sigma_{j}^{2}\alpha_{j}\beta_{j}+\beta_{j}(3-2\sigma_{j}^{2}\beta_{j})]+4\bar{\mu}_{j}[\alpha_{j}-2\sigma_{j}^{2}\alpha_{j}\beta_{j}+\beta_{j}(-5+2\sigma_{j}^{2}\beta_{j})]\nonumber\\
&+\beta_{j}[\beta_{j}(2\beta_{j}(1+2\sigma_{j}^{2}(\beta_{j}-\bar{\mu}_{j}))+2\bar{\mu}_{j})+2\alpha_{j}(3\bar{\mu}_{j}+\beta_{j}(-5-2\sigma_{j}^{2}(\beta_{j}-\bar{\mu}_{j})))]\Big]
\text{erfc} \left(\sqrt{\sigma_{j}^{2}\beta_{j}}\right) \Bigg\}\nonumber\\
&+\frac{4\sqrt{2/\pi}}{(\beta_{j}-\alpha_{j})^{3}}\Bigg\{2\sigma_{j}(\alpha_{j}-\beta_{j})[\alpha_{j}^{2}+\beta_{j}^{2}-2\bar{\mu}_{j}(\alpha_{j}+\beta_{j})]+\sqrt{\pi}\sqrt{\alpha_{j}}e^{\sigma_{j}^{2}\alpha_{j}}\Big[2\sigma_{j}^{2}\alpha_{j}^{3}+\alpha_{j}^{2}[1-2\sigma_{j}^{2}(\beta_{j}+2\bar{\mu}_{j})]+6\beta_{j}\bar{\mu}_{j}\nonumber\\
&+\alpha_{j}[-5\beta_{j}+2\bar{\mu}_{j}+4\sigma_{j}^{2}\beta_{j}\bar{\mu}_{j}]\Big]\Big[-1+erf(\sqrt{\sigma_{j}^{2}\alpha_{j}})\Big]
+\sqrt{\pi}\sqrt{\beta_{j}}e^{\sigma_{j}^{2}\beta_{j}}\Big[\beta_{j}[\beta_{j}+2\sigma_{j}^{2}\beta_{j}(\beta_{j}-2\bar{\mu}_{j})+2\bar{\mu}_{j}] \nonumber\\
&+\alpha_{j}[6\bar{\mu}_{j}+\beta_{j}(-5-2\sigma_{j}^{2}(\beta_{j}-2\bar{\mu}_{j}))]\Big]\text{erfc} \left(\sqrt{\sigma_{j}^{2}\beta_{j}}\right)\Bigg\}. \nonumber
\end{align}
Again in the limit $\sigma\rightarrow 0 $, the disorder corrections to the EoS $h_j$ reduces to that obtained for the delta-correlated disorder \cite{Boudj200}. 
For $\sigma\rightarrow 0 $ and $\Delta\rightarrow\infty$ (or $g_{12} \rightarrow 0$, equivalently), one has $h_{j}(\infty,0)=6$
hence, the EoS (\ref{GF2}) simplifies to that of a single component BEC namely : $\mu = g n(1+12\pi R' \sqrt{n a^3/\pi})$ found in Refs.\cite{Falco,Gior,Yuk}.

\begin{figure}[htb] 
\includegraphics[scale=0.7]{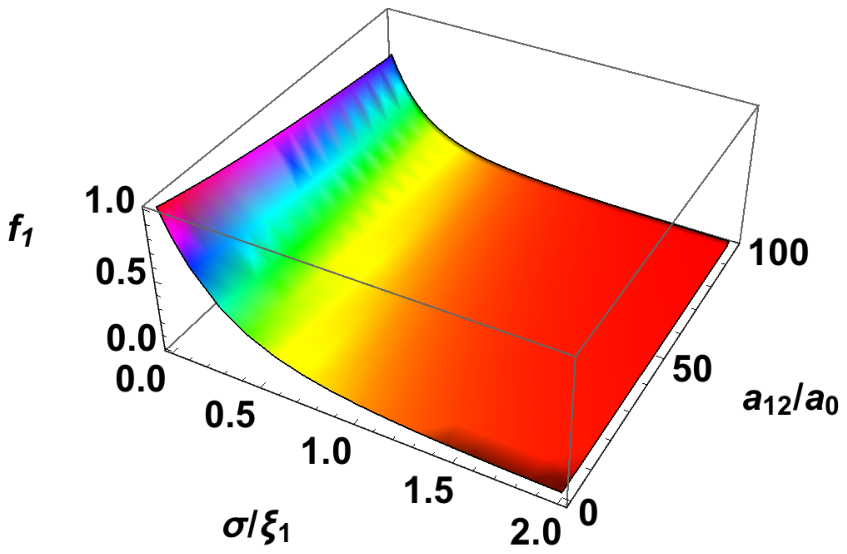}
\includegraphics[scale=0.7]{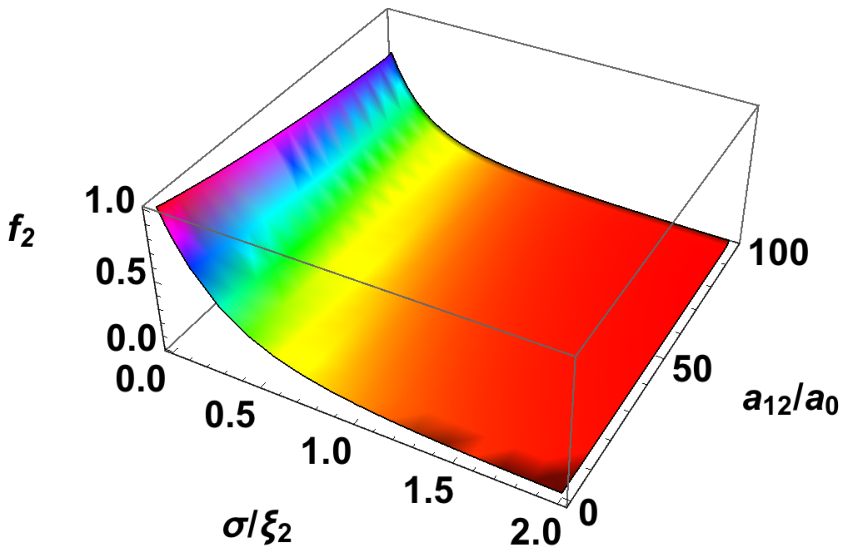}
\caption {(Color online)  Behavior of the disorder functions $f_j$ as a function of the interspecies interaction strength $a_{12}$ and the disorder correlation length $\sigma$ 
for ${}^{87}$Rb-${}^{87}$Rb mixture in two different internal states. 
Parameters are:  $a_1= 100.4\,a_0$ and $a_2= 95.44\,a_0$ \cite{Ego},
and the densities:  $n_1=n_2=10^{21}$m$^{-3}$.}
\label{dis}
\end{figure}

\begin{figure}[htb] 
\includegraphics[scale=0.8]{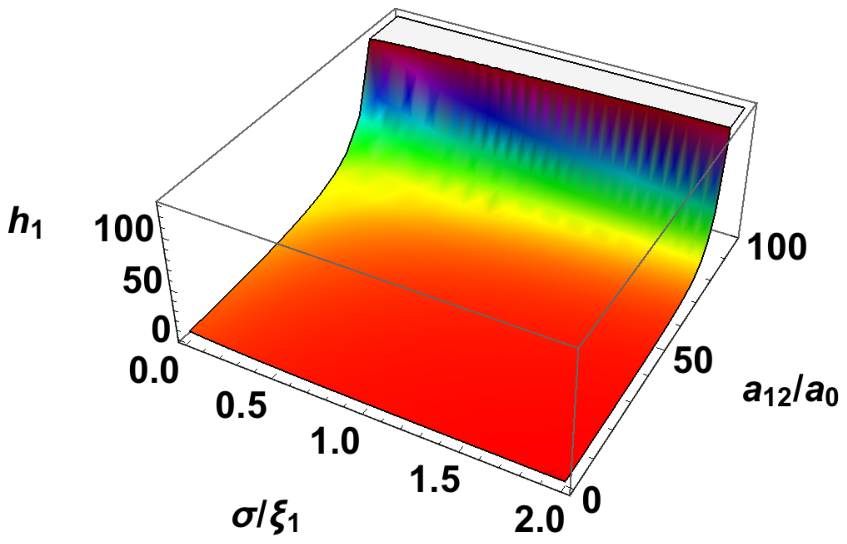}
\includegraphics[scale=0.8]{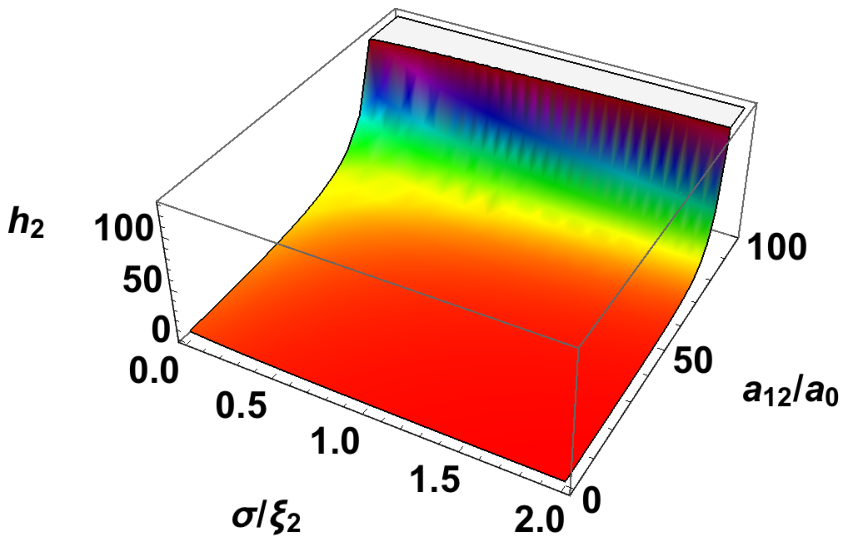}
\caption {(Color online) Behavior of the disorder functions $h_j$ as a function of the interspecies interaction strength $a_{12}$ and the disorder correlation length 
for ${}^{87}$Rb-${}^{87}$Rb mixture.
Parameters are the same as in Fig.\ref{dis}.}
\label{dis11}
\end{figure}

Figure \ref{dis11} depicts that functions $h_j$ are increasing with the interspecies interactions strength $a_{12}$
results in an enhacement of the total chemical potential.

\end{widetext}

\end{document}